\documentclass[superscriptaddress,twocolumn,showpacs,preprintnumbers,amsmath,amssymb,prb]{revtex4}

\usepackage{graphicx}
\usepackage{dcolumn}
\usepackage{bm}

\begin{document} 

\title{Vortex lattice structure in BaFe$_2$(As$_{0.67}$P$_{0.33}$)$_2$ by the small-angle neutron scattering technique}

\author{R. Morisaki-Ishii}
\affiliation{Division of Natural/Applied Science, Graduate School of Humanities and Science, Ochanomizu University, Bunkyo-ku, Tokyo 112-8610, Japan}
\author{H. Kawano-Furukawa}
\affiliation{Division of Natural/Applied Science, Graduate School of Humanities and Science, Ochanomizu University, Bunkyo-ku, Tokyo 112-8610, Japan}
\author{A.S. Cameron}
\affiliation{School of Physics and Astronomy, University of Birmingham B15 2TT, United Kingdom}
\author{L. Lemberger}
\affiliation{School of Physics and Astronomy, University of Birmingham B15 2TT, United Kingdom}
\author{E. Blackburn}
\affiliation{School of Physics and Astronomy, University of Birmingham B15 2TT, United Kingdom}
\author{A.T. Holmes}
\affiliation{School of Physics and Astronomy, University of Birmingham B15 2TT, United Kingdom}
\author{E.M. Forgan}
\affiliation{School of Physics and Astronomy, University of Birmingham B15 2TT, United Kingdom}
\author{L.M. DeBeer-Schmitt}
\affiliation{Neutron Sciences Directorate, Oak Ridge National Laboratory, Oak Ridge, TN 37831, USA}
\author{K. Littrell}
\affiliation{Neutron Sciences Directorate, Oak Ridge National Laboratory, Oak Ridge, TN 37831, USA}
\author{M. Nakajima}
\affiliation{Department of Physics, University of Tokyo, 7-3-1 Hongo, Bunkyo-ku, Tokyo 113-0033, Japan}
\author{K. Kihou}
\affiliation{National Institute of Advance Industrial Science and Technology (AIST), Tsukuba, Ibaraki 305-8568, Japan}
\author{C.H. Lee}
\affiliation{National Institute of Advance Industrial Science and Technology (AIST), Tsukuba, Ibaraki 305-8568, Japan}
\author{A. Iyo}
\affiliation{National Institute of Advance Industrial Science and Technology (AIST), Tsukuba, Ibaraki 305-8568, Japan}
\author{H. Eisaki}
\affiliation{National Institute of Advance Industrial Science and Technology (AIST), Tsukuba, Ibaraki 305-8568, Japan}
\author{S. Uchida}
\affiliation{Department of Physics, University of Tokyo, 7-3-1 Hongo, Bunkyo-ku, Tokyo 113-0033, Japan}
\author{J.S. White}
\affiliation{Laboratory for Neutron Scattering, Paul Scherrer Institut, CH-5232 Villigen, Switzerland}
\affiliation{Laboratory for Quantum Magnetism, Ecole Polytechnique F$\acute{e}$d$\acute{e}$rale de Lausanne (EPFL), CH-1015 Lausanne, Switzerland}
\author{C.D. Dewhurst}
\affiliation{Institut Laue-Langevin, Grenoble 38042, France}
\author{J.L. Gavilano}
\affiliation{Laboratory for Neutron Scattering, Paul Scherrer Institut, CH-5232 Villigen, Switzerland}
\author{M. Zolliker}
\affiliation{Laboratory for Developments and Methods, Paul Scherrer Institute, CH-5232 Villigen, Switzerland}
\date{\today}
\begin{abstract}
{We have observed a magnetic vortex lattice (VL) in BaFe$_2$(As$_{0.67}$P$_{0.33}$)$_2$ (BFAP) single crystals by small-angle neutron scattering (SANS). With the field along the {$\bf c$}-axis, a nearly isotropic hexagonal VL was formed in the field range  from 1 to 16 T, which is a record for this technique in the pnictides, and no symmetry changes in the VL were observed. The temperature-dependence of the VL signal was measured and confirms the presence of (non $d$-wave) nodes in the superconducting gap structure for measurements at 5 T and below. The nodal effects were suppressed at high fields. At low fields, a VL reorientation transition was observed between 1 T and 3 T, with the VL orientation changing by $45^{\circ}$. Below 1 T, the VL structure was strongly affected by pinning and the diffraction pattern had a fourfold symmetry. We suggest that this (and possibly also the VL reorientation) is due to pinning to defects aligned with the crystal structure, rather than being intrinsic.}
\end{abstract}

\pacs{74.25.Uv, 74.25.Dw, 74.70.Xa}

\maketitle{}

\section{INTRODUCTION}
BaFe$_2$(As$_{0.67}$P$_{0.33}$)$_2$ is one of the most well-researched materials among the 122-family of iron-pnictide superconductors because of its tremendous possibilities to provide an answer about the systematic interpretation of the gap structure and superconducting pairing symmetry. BFAP is a favorable material to investigate, because single crystals can be prepared with high enough quality to observe quantum oscillations - the de Haas-van Alphen effect\cite{Shishido}. In consequence, the nature of superconductivity in BFAP is becoming clearer, as described below. The possibility that BFAP has line nodes in the superconducting order parameter is indicated by various experiments such as penetration depth\cite{Hashimoto}, thermal conductivity\cite{Hashimoto,Yamashita}, and nuclear magnetic resonance (NMR) relaxation rate measurements\cite{Nakai} on single crystals of BFAP. Moreover, the existence of nodes in iron pnictides with $s$-wave pairing symmetry is also suggested by angle-resolved photoemission (ARPES) measurements\cite{Zhang}. On the other hand, a spin resonance in an $s_{\pm}$-wave iron-based superconductor without line nodes has recently been observed by inelastic neutron scattering measurements\cite{Ishikado}. This requires a sign-change in the order parameter so suggests $s_{\pm}$ pairing. In addition, theoretical models using first-principles calculations suggest that BFAP has a three-dimensional nodal structure with a $s_{\pm}$-wave gap originating from the spin fluctuations\cite{Suzuki}. For a comprehensive understanding of the superconducting state in BFAP, it is valuable to detect the magnetic vortex lattice directly by small-angle neutron scattering experiments.   

 SANS measurements on superconducting VL states can give important information on the electron pairing state. This technique can be used to explore the vortex state, since it measures the Fourier components of the magnetic field distribution in a crystal. SANS is also a powerful technique to observe directly the VL geometry in a bulk crystal.
So far, just three pnictide materials are reported to show VL signals. However, in Ba(Fe$_{0.93}$Co$_{0.07}$)$_{2}$As$_2$ (BFCA) \cite{Eskildsen} and LiFeAs\cite{Inosov} single crystals, only ring-like diffraction patterns were observed. Recently, KFe$_2$As$_2$ (KFA) single crystals showed clear magnetic VL Bragg spots for the first time in iron-based superconductors\cite{Furukawa, ROPP}. In KFA, a nearly isotropic hexagonal VL is formed, and no symmetry transitions up to 2/3 $B_{\rm {c2}}(T=0)$ are observed. From these SANS experiments, it was concluded that the gap in KFA  has line nodes with s$_{+-}$-symmetry; these nodes may be ``horizontal", running around Fermi surfaces which have cylindrical topology, or have a more complicated structure\cite{Hashimoto3,Sato}. 
In order to further expose systematically the nature of iron-based superconductors especially for the 122-type structure, it is of importance to explore more systems that show a clear VL signal. Therefore, we have performed SANS experiments on large high quality single crystals of BFAP. We have determined whether, similar to KFA, line nodes are present.

\section{EXPERIMENTAL DETAILS}
\begin{figure}[htbp]
\begin{center}
 \includegraphics [width=85mm]{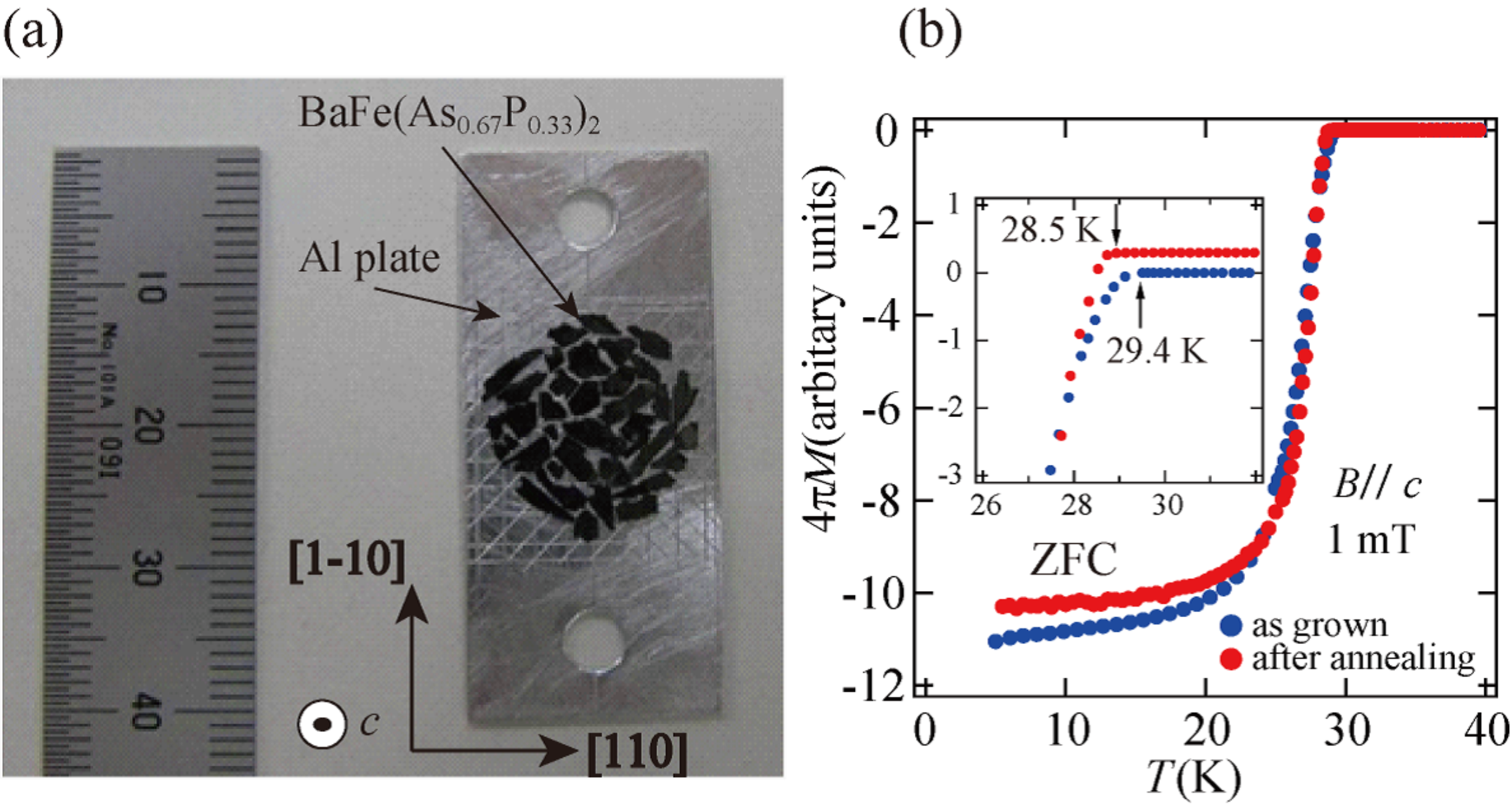}
\end{center}
\caption{(color online) (a) BFAP single crystals on an aluminum plate which we used in the SANS experiments. (b) The temperature dependence of the susceptibility measured at $\mu_0H = 1$~mT for ${\bf B}\parallel {\bf c}$. Data for the as-grown sample and annealed sample are shown by the blue and red circles, respectively. Inset: enlargement of main panel around $T_{\rm c}$. The corresponding values of $T_{\rm c}$ are shown next to the arrows. For visibility in the inset, $M$ of the annealed crystal is shifted upward by 0.3 units.}
\label{sample}
\end{figure}

The SANS experiments were carried out on the D11 and D33 instruments at the Institut Laue-Langevin (ILL), France, the SANS-I instrument, Swiss Spallation Neutron Source (SINQ) at the Paul Scherrer Institute (PSI), Switzerland, and the CG2 instrument at the Oak Ridge National Laboratory (ORNL), USA. Neutrons with a wavelength of $\lambda_n$ = 5$\sim$9 \AA \ with $\delta\lambda_n$/$\lambda_n$ = 10\% or 15\%, were used, depending on the instrumental situation. We measured VL scattering patterns in the ($hk0$) plane with various magnetic fields  up to 16 T applied parallel to the {$\bf c$}-axis and approximately parallel to the incident beam of neutrons. 
For the present study, single crystals of optimally-doped BFAP with $T_{\rm c}$ $\sim$ 30 K were grown by the self-flux method\cite{Nakajima}. The typical size of a crystal was $\sim$ 3 $\times$ 3 $\times$ 0.1 mm$^3$, and we used about 200 single crystals with a total mass of $\sim$ 0.28 g. These were aligned side by side and attached with CYTOP$^{\textregistered}$ hydrogen-free glue onto aluminum plates. One of the plates which we used in the experiments is shown in Fig.\ \ref{sample}(a). We annealed the crystals in vacuum at 500\ $^\circ$~C for 20 hours just before the SANS experiments, with the intention of improving the perfection of the crystals. In a previous SANS experiment on BFAP crystals without this treatment, we saw only a ring-shaped signal due to vortex pinning by defects. We performed susceptibility measurements using a piece of BFAP single crystal for checking the sample quality before and after the annealing process.  Under zero-field-cooled conditions, with magnetic field $\mu_0H$ = 1 mT applied along the {\bf c}-axis, we found that the magnetization ($M$) changes rapidly below the onset temperature $T_{\rm c}$ = 30 K of the superconducting transition, which agrees with previous reports\cite{Kasahara,Chong}. After annealing, the maximum ZFC magnetization below 20 K increased, $T_{\rm c}$ was slightly lower, and $M$ changed more rapidly just below $T_{\rm c}$ (see Fig.\ \ref{sample}(b) and its inset). This is because the lattice defects which lead to pinning are reduced. From these results, it is clear that after annealing the sample had better crystallinity than as-grown.  

SANS measurements at various fields were performed on these annealed BFAP single crystals, and we have observed clear diffraction signals from a well-ordered magnetic VL in BFAP up to 16 T, which is a record for the pnictides. This is the second success in observing VL Bragg spots among the iron-based superconductors following on from KFA. Here, we report the temperature and magnetic field dependences of the SANS patterns and of the integrated intensity. The field was applied parallel to the {\bf c}-axis above $T_c$ and oscillated in value by $\pm$ 0.05 T  as the sample was cooled to base temperature, and then held constant. We used a scattering geometry with the field nearly parallel to the neutron beam. Data were collected as the field and sample were rotated together through a set of angles, so that VL spots were ``rocked" through the Bragg condition. Background data collected at 35 K were subtracted from the data taken at 2 K. The SANS data were displayed and analyzed by the GRASP analysis software\cite{GRASP}.  The scattering patterns displayed in this paper were prepared by Bayesian treatment of the data\cite{Alex}. In this analysis based on Bayes' theorem, a weighted average of data is used to emphasise data from regions of the detector satisfying the Bragg condition for each VL angle. 

\section{EXPERIMENTAL RESULTS}
\subsection{Vortex lattice structure}

\begin{figure*}[!ht]
\begin{center}
 \includegraphics [width=140mm]{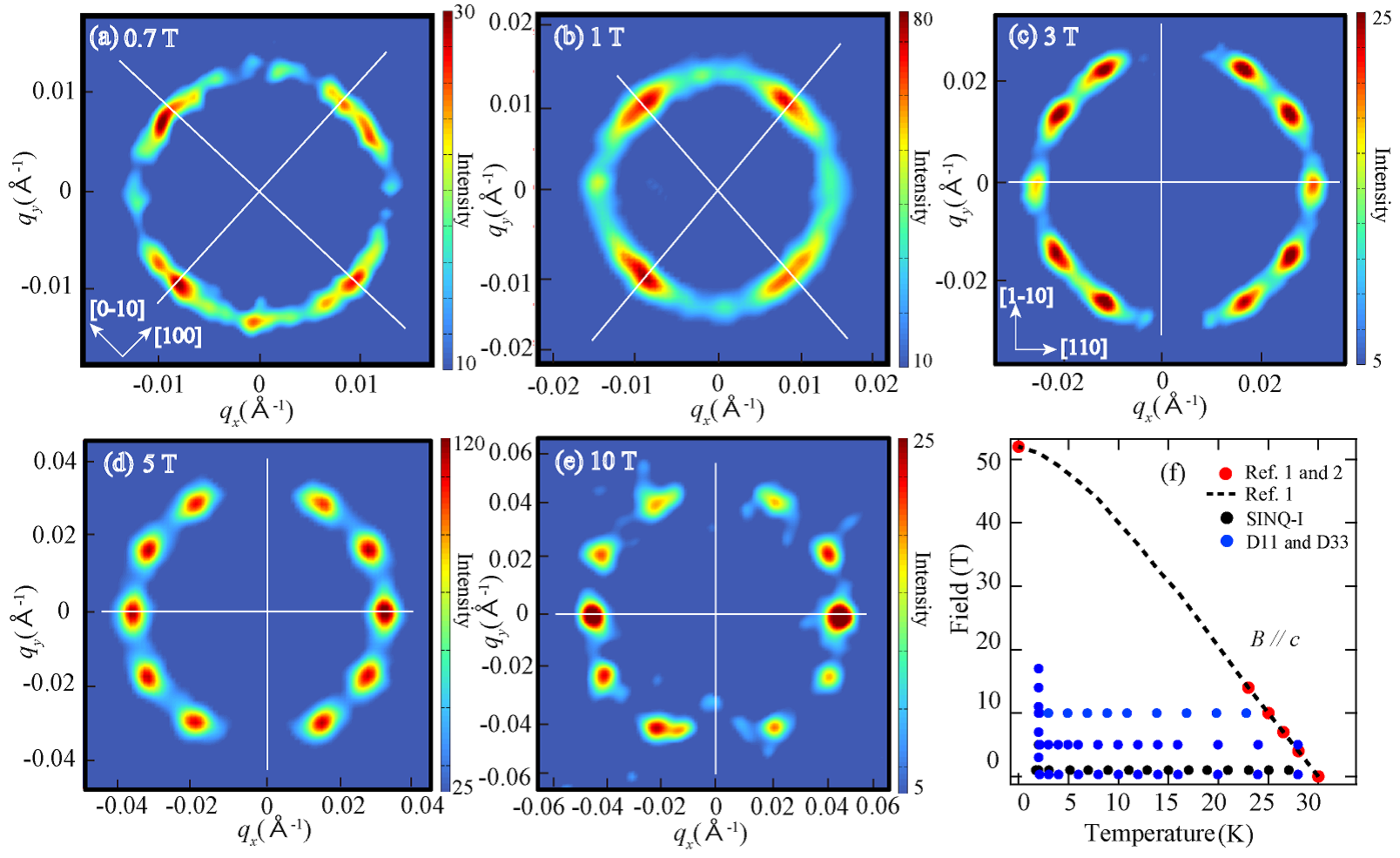}
\end{center}
\caption{(color online) Typical small-angle neutron scattering patterns at 2 K from the VL structure in BaFe$_2$(As$_{0.67}$P$_{0.33}$)$_2$ at (a) 0.7 T, (b) 1 T, (c) 3 T, (d) 5 T, and (e) 10 T. The $\langle 110 \rangle$ directions of the crystals are vertical and horizontal. Mirror planes for the VL are shown by the white lines in each figure. After the Bayesian analysis, to improve visibility, the data were smoothed with Gaussian of width comparable to the instrumental resolution, and Poisson noise near the direct beam was masked. The color scale from this analysis represents the integrated intensity per pixel in arbitrary units. (f) The $B$-$T$ phase diagram indicating all conditions under which VL measurements were performed. The values of $B_{\rm {c2}}(T)$ were taken from Refs.~\onlinecite{Shishido} and \onlinecite{Hashimoto}.}
\label{VL}
\end{figure*}

\begin{figure}[htbp]
\begin{center}
 \includegraphics [width=70mm]{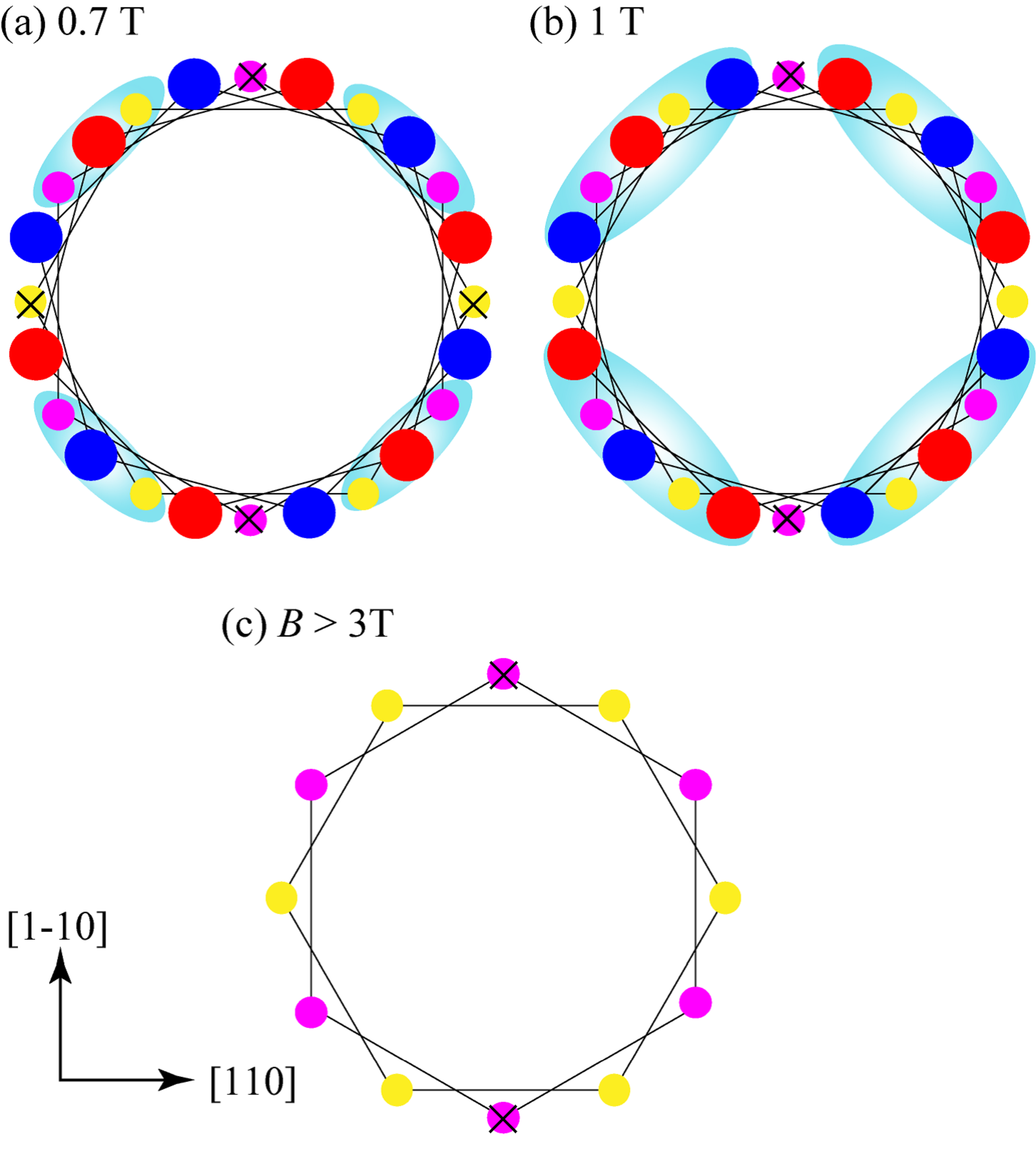}
\end{center}
\caption{(color online) (a) (b) Schematic diffraction patterns from four types of hexagonal VLs. The resultant VL diffraction patterns are highlighted by light blue ellipses. Some spots with $\times$ were not observed in our experiments. (c) Hexagonal VLs with two domain orientations, observed at high field.}
\label{Sketch}
\end{figure}
Figures \ref{VL}(a)-(e) show the VL diffraction patterns with clear Bragg spots measured at 2 K in selected magnetic fields along the {\bf c}-axis. Mirror axes of the VL structures are shown by the white lines on the VL scattering patterns. Note that the \emph{real space} VL nearest-neighbors are in the same pattern as the first-order diffraction spots around the direct beam, but rotated $90^{\circ}$ around the field axis. 

We observed ring-like patterns with Poisson noise at low fields of 0.05 T to 0.4 T, with some concentration of the intensity along the $\langle 1\ 0\ 0 \rangle$ directions. On increasing the field, Bragg spots emerge at $\sim0.7$ T (Fig. 2(a)) and we see a clear 4-spot pattern along the $\langle 1\ 0\ 0 \rangle$ directions at 1 T (Fig. 2(b)). At first glance, it appears that the vortices form a square VL; however, a more careful inspection shows weak signals in both the horizontal and vertical directions. We believe this means that the 4-spot pattern observed at 1 T actually indicates not the formation of square VL, but instead the presence of two distorted hexagonal VL domains. Each VL would have two strong spots aligned along a diagonal, with weak signals in the horizontal and vertical directions arising from the other spots. This conclusion is strengthened by the experimental value of the average wave vector $q_{\rm exp}$ = 0.0147(1) \AA $^{-1}$, which is much closer to the expected hexagonal value of $q_{\rm hex.}$ = 0.0148 \AA $^{-1}$ at 1 T than the square value, $q_{\rm sq.}$ = 0.0138 \AA $^{-1}$. On applying fields above 1 T, the spots begin to split and slightly distorted hexagonal VLs appear at 3 T (Fig. 2(c)). (Note that when these higher-field data were taken, scans through vertical spots were not experimentally possible, and so these spots, which must be present by symmetry, were not measured.) Finally, 12 clear spots would be present in a full diffraction pattern from 5 T upwards, of which 10 were observed. The 12 spots are derived from a mixture of two domain orientations of hexagonal VLs and are present in the range of at least 3 T to 16 T. Because these clear Bragg spots were observed up to  more than 20~K for 5 and 10 T, we may conclude that a hexagonal VL is stable throughout the superconducting mixed state above 5 T at least. From these results, we find that BFAP is the first example of a non-stoichiometric iron-based superconductor which shows well-ordered VLs. This may be a consequence of isovalent doping on the As site, instead of non-isovalent doping on the Fe site as in Co-doped pnictide Ba(Fe$_{0.93}$Co$_{0.07}$)$_2$As$_2$.\cite{Eskildsen}

\begin{figure}[htbp]
\begin{center}
 \includegraphics [width=80mm]{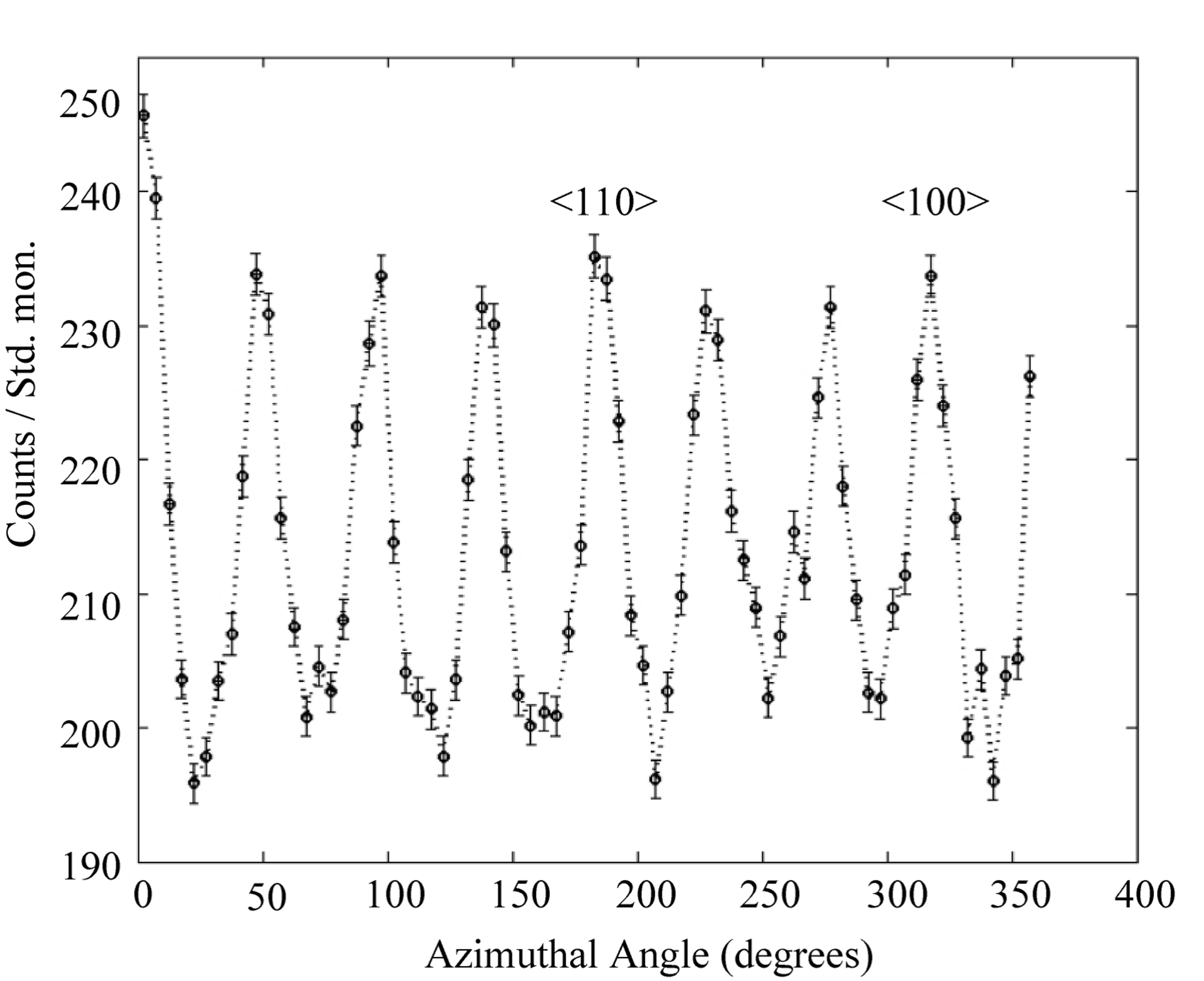}
\end{center}
\caption{The angular variation of background scattering from the sample. This has streaks on an isotropic background. They are multiples of $45^{\circ}$ apart and lie along $\langle 1\ 0\ 0 \rangle$ and  $\langle 1\ 1\ 0 \rangle$  directions (two typical directions are labeled).}
\label{bkg}
\end{figure}

Here, we consider the $45^{\circ}$ rotation of the hexagonal VL in the $ab$-plane. As noted above, the main signals from a hexagonal VL are seen along the $\langle 1\ 0\ 0 \rangle$ directions at 0.7 T and 1 T, and the mirror axes of the VL structure lie along these directions. In contrast, the directions of the VL mirror planes above 3 T are rotated by $45^{\circ}$, compared with those at 0.7 T and 1 T. It is possible that this is due to a first-order transition, with the VL rotating by $45^{\circ}$ between 1 T and 3 T. With consideration of these results, the hexagonal VLs drawn in Figs. \ref{Sketch}(a)-(c) may account for the diffraction patterns observed in our SANS measurements. We suggest that the $45^{\circ}$ structure is mainly dominant, but the hexagonal VLs which are realized above 3 T are partly present between 1 \& 3 T. At the lowest fields, the pattern remains ring-like, but with extra intensity along the $\langle 1\ 0\ 0 \rangle$ directions. In the region of $q$-values appropriate to fields below 1 T, we find extra non-VL scattering from the sample along both  $\langle 1\ 0\ 0 \rangle$ and  $\langle 1\ 1\ 0 \rangle$ directions: see  Fig. \ref{bkg}. This will arise from planar crystal defects, which are likely to have similar spacing to the vortex lines at these fields. It appears that the $\langle 1\ 0\ 0 \rangle$ defects are particularly effective in controlling the orientation of pinned VL planes. In this field region, the VL rocking curves become much broader, indicating that the \emph{directions} of the vortex lines are distorted by pinning away from parallel to the field. We also find that the $q$-vector distribution in the diffraction patterns becomes broad, with a peak close to that expected for a square VL along the $\langle 1\ 0\ 0 \rangle$ directions. However, along the $\langle 1\ 1\ 0 \rangle$ directions, the peak is close to that for a hexagonal VL, rather than the $\sqrt2 \times$ larger value expected for higher-order reflections from an intrinsic square VL. We conclude that in some regions of the sample, crossing planar defects are sufficiently similar in density to the vortex lines to control the local VL coordination at low fields; however, in other regions, an intrinsic hexagonal VL remains. However at fields above 3 T, the situation appears more straightforward, with a mixture of two equivalent domain orientations of hexagonal VLs, as drawn in Fig. \ref{Sketch}(c). 

\subsection{Vortex lattice phase diagram}
\begin{figure}[ht]
\begin{center}
 \includegraphics [width=80mm]{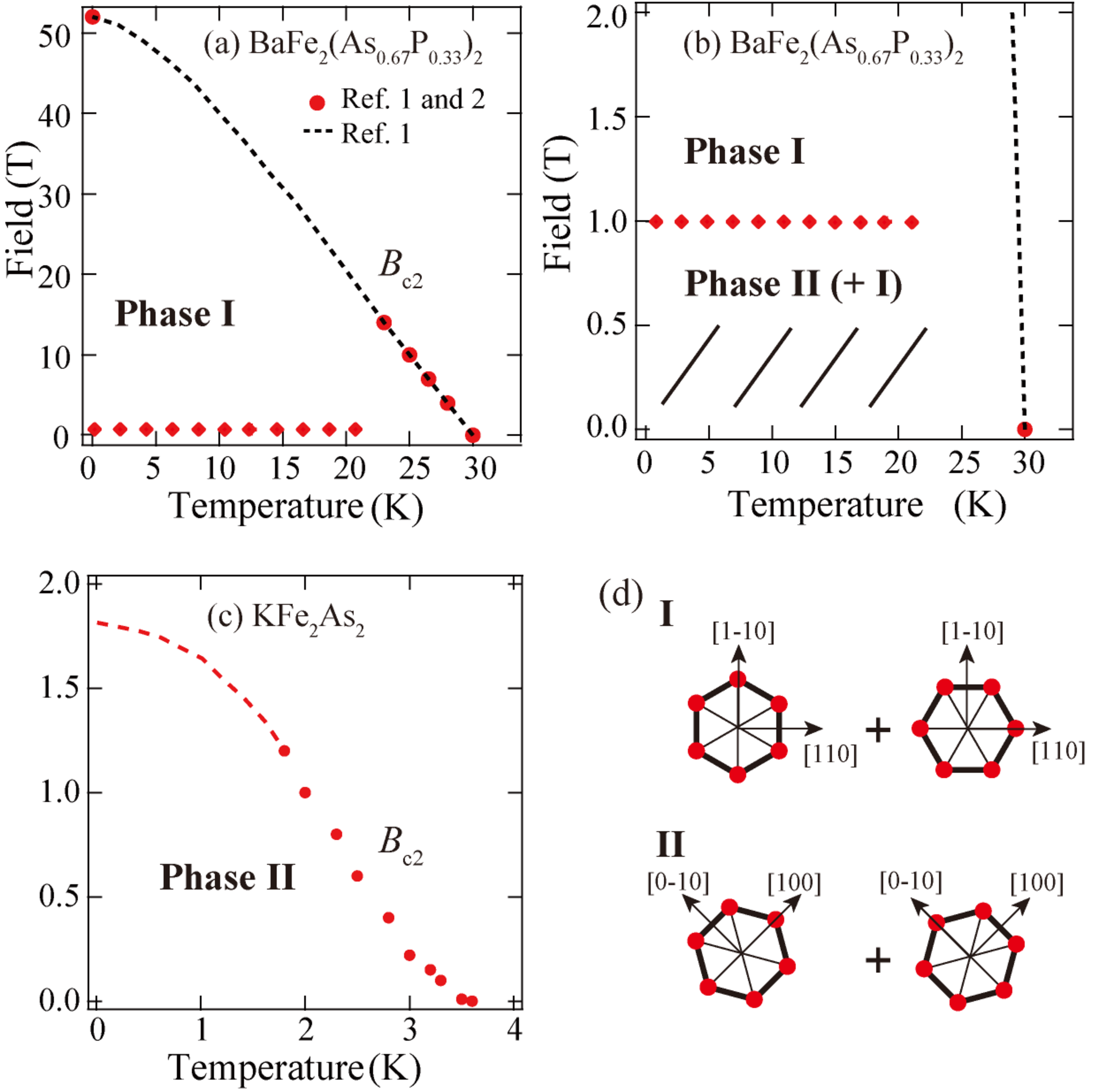}
\end{center}
\caption{(color online) (a) Schematic $B-T$ phase diagrams of VL structure for BaFe$_2$(As$_{0.67}$P$_{0.33}$)$_2$. Red dots and a black dashed line (from Refs. \onlinecite{Shishido} and \onlinecite{Hashimoto}) give a rough indication of $B_{\rm {c2}}$ for BFAP. (b) The low-field region below 1 T in the VL phase diagram for BFAP, showing the pinning-dominated region as a hatched area. (c) Schematic $B-T$ phase diagrams of VL structure for KFe$_2$As$_2$.\cite{Furukawa} Red dots and a red dashed line (from Ref.~\onlinecite{Furukawa}) provide estimates of $B_{\rm {c2}}$ for KFA.  On increasing the temperature, hexagonal VL structures are stable up to $T_{c2}(B)$ for both BFAP and KFA. However, the hexagonal lattice in Phase I for BFAP is rotated by $45^{\circ}$, compared to that of Phase I\hspace{-.1em}I in BFAP, which is the only phase observed in KFA. (d) Diagrams of the VL structures which are proposed for Phase I and I\hspace{-.1em}I. }
\label{HT}
\end{figure}

\begin{figure}[htbp]
\begin{center}
 \includegraphics [width=70mm]{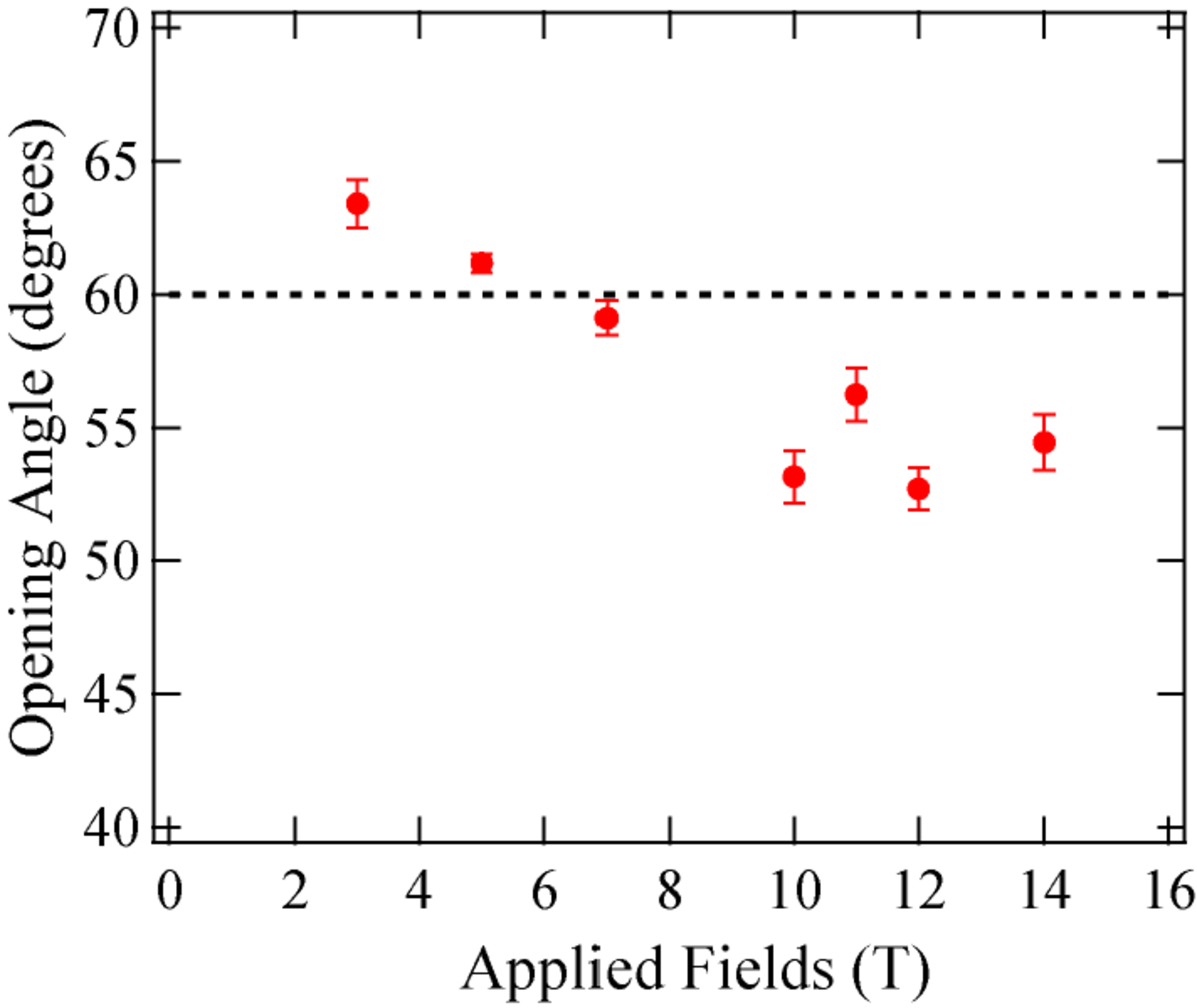}
\end{center}
\caption{The variation of the VL shape with field, represented by the  angle between the VL spots which is bisected by the vertical and horizontal  $\langle 1\ 1\ 0 \rangle$  directions. This is close to the regular hexagonal value of $60^{\circ}$ at all fields where it could be established accurately.}
\label{distn}
\end{figure}

A schematic VL $B$-$T$ phase diagram of the VL structure for BFAP is drawn in Fig. \ref{HT}(a), and the low-field region below 1 T is shown in Fig. \ref{HT}(b). For comparison, VL phase diagram of KFA is shown in Fig. \ref{HT}(c). The orientations of the VLs to the crystal axes for BFAP and KFA are shown in Fig. \ref{HT}(d). A hexagonal VL structure is dominant in BFAP as well as in $s$-wave superconductor KFA, where no symmetry transition is observed up to 2/3$B_{\rm c2}(T=0)$.~\cite{Furukawa} From the present results, the possibility of $d$-wave superconductivity in BFAP can also be ruled out, because a square VL was not observed in BFAP at high fields, in contrast with the typical $d$-wave superconductor of CeCoIn$_5$.\cite{Bianchi,Kawamura} On the other hand, at low fields there may exist an extra VL phase in addition to Phase I: the $45^{\circ}$ rotated  Phase I\hspace{-.1em}I, together with pinning-dominated behaviour at low fields. The VL reorientation is very like the $45^{\circ}$ rotation seen in borocarbide superconductors {\it R}Ni$_2$B$_2$C ({\it R} = Y, Rare earth).\cite{Sakiyama,Furukawa2} This arises from nonlocal effects\cite{Kogan,N_Nakai}, which should also give rise to a square VL at higher fields, as seen in experiments in borocarbides. However, no high-field VL transitions are seen in BFAP; this suggests that the VL reorientation in our crystals is not due to nonlocal effects. Further evidence for the weakness of nonlocal effects (and the absence of a $d$-wave order parameter) in BFAP is given in Fig. \ref{distn}, which shows that the VL is very close to regular hexagonal at all fields. Because the VL structure at the lowest fields appears to be dominated by pinning to crystal defects, it is possible that the reorientation transition is also a relic of this rather than being intrinsic.

\subsection{Field dependence of form factor}
\begin{figure}[htbp]
\begin{center}
 \includegraphics [width=70mm]{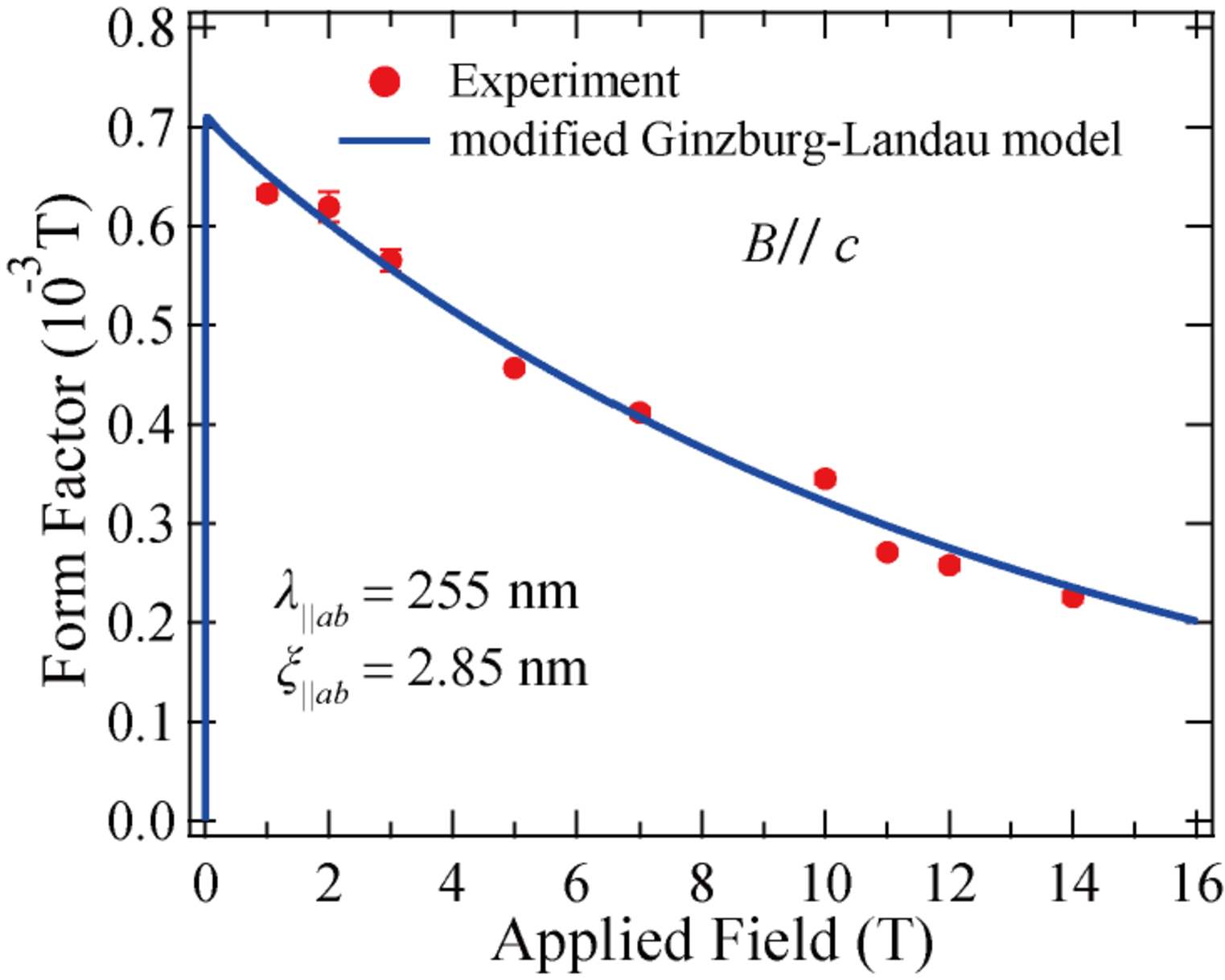}
\end{center}
\caption{(color online) Field dependence of the form factor ($F$) at 2 K. Experimental data are fitted by the modified London model. }
\label{formfactor} 
\end{figure}

Figure \ref{formfactor} shows the field dependence of the VL form factor, derived from the intensity $I_{\bf q}$ of the VL first-order diffraction spots at 2 K, after integration over the rocking angle.
As derived in Ref.~\onlinecite{Christen}, $I(q)$ is related to the form factor by:
\begin{eqnarray}
I_{\bf q}=2\pi V \phi \Big(\frac{\gamma}{4}\Big)^2\frac{\lambda_n^2}{\Phi_0^2 q}|F({\bf q})|^2,
\label{IntIEqn}
\end{eqnarray}
where $F({\bf q})$ is the VL form factor, $V$ is the sample volume, $\phi$ is the incident neutron flux, $\gamma$ = 1.91 is the neutron gyromagnetic ratio, and $\Phi_0$ = 2.07$\times$10$^{-15}$ T is the flux quantum.  $F({\bf q})$ is a measure of the magnitude of the spatial variation of the field in the mixed state. 
These data were fitted to a modified London model with core/nonlocal corrections\cite{Ichioka}, 
\begin{eqnarray}
F(q)=\frac{B}{1+q^2\lambda^2}\exp(-cq^2\xi^2),
\label{FFEqn}
\end{eqnarray}
where, $\lambda$ is the London penetration depth, $\xi$ is the coherence length, and $c$ is an empirical core cutoff parameter that might range from 1/4 to 2.\cite{Yaouanc} Here, we adopt an appropriate value of 0.44 for $c$.\cite{Bowell,White} The best fit gave $\lambda_{\parallel {ab}}$ = 255 $\pm$ 3 nm and $\xi_{\parallel {ab}}$ = 2.85 $\pm$ 0.07 nm. From this, the value of the critical field  at $T$ = 2~K is found to be 40.6 T by the relationship $B_{\rm {c2}}^c = \Phi_0 / \pi \xi_{\parallel {ab}}^{2}$. This value is close to that of derived by the extrapolated critical field $B_{\rm{c2}}$ = 52 T from torque measurements \cite{Hashimoto,Shishido}. The small difference may be the result of minor distortions of the VL by pinning, which can lead to a ``static Debye-Waller factor" which mimics a larger core size $\xi$. Our derived value of $\lambda$ is larger than, but comparable to the estimated value of 200 $\pm$ 30 nm from the low-$T$ magnetic penetration depth measurement\cite{Hashimoto}. The Ginzburg-Landau (GL) parameter derived from our results is $\kappa$ $\sim$ $\lambda$/$\xi$ $\sim$ 89.5 for {\bf B} $\parallel$ {\bf c}.

\subsection{Temperature-dependence of Intensity}
\begin{figure}[htbp]
\begin{center}
 \includegraphics [width=85mm]{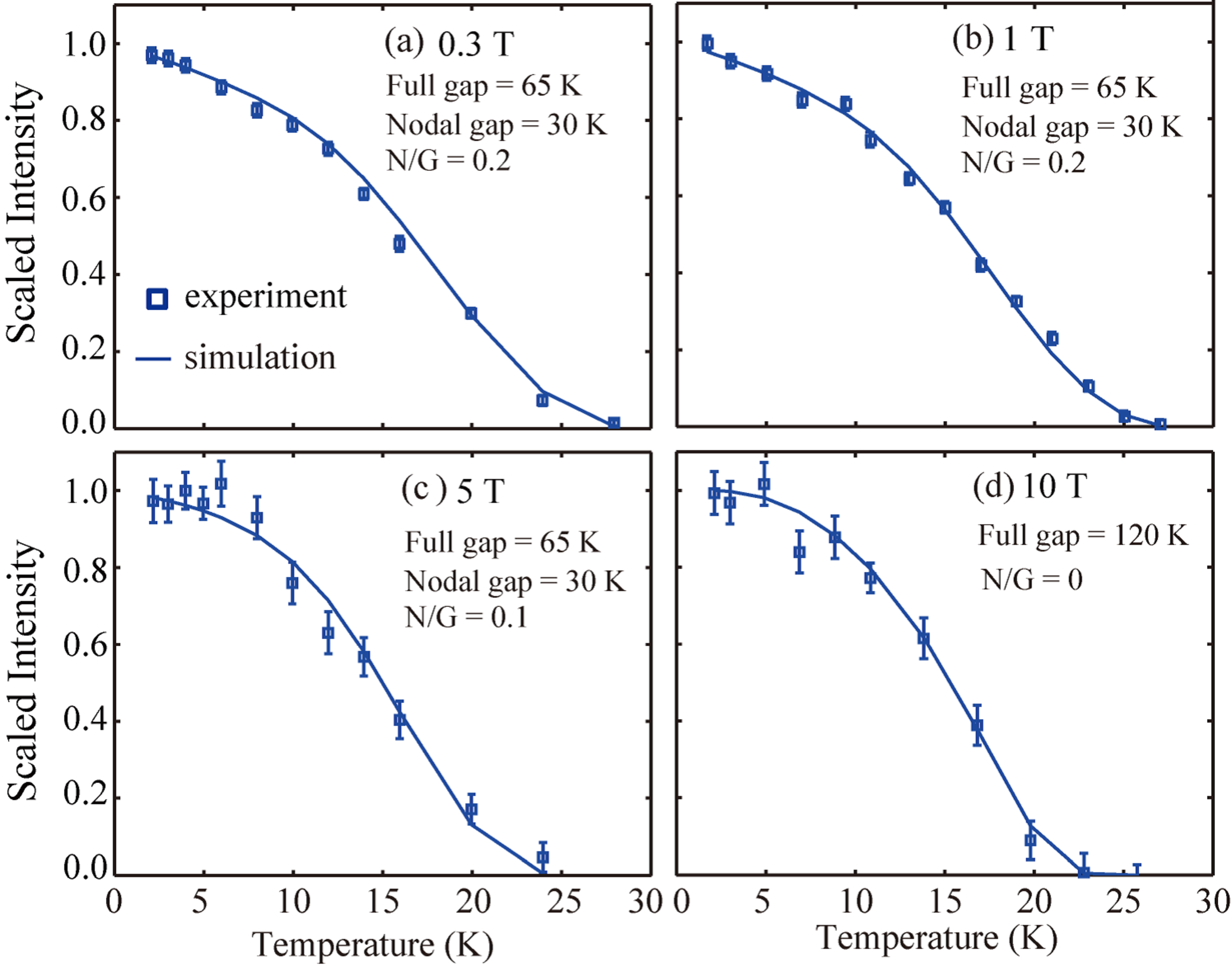}
\end{center}
\caption{(color online) Experimental data for the scaled intensity $I$ $vs.$ $T$(K) at selected fields. We show the fits to the experimental data, the relative weight of nodal gap and estimated gap values at (a) 0.3 T, (b) 1 T, (c) 5 T, and (d) 10 T. }
\label{IT}
\end{figure}

We show the temperature dependence of the scaled intensity at 0.3 T, 1 T, 5 T, and 10 T in Fig. \ref{IT}. We see that the linear low-$T$ behavior tends to disappear as large fields are applied.

To fit data at a fixed field as a function of temperature to a model, we need to calculate  the form factor using Eq. (\ref{FFEqn}), to give the integrated intensity, which is proportional to $F^2$ (Eq. (\ref{IntIEqn})) . The most important temperature-dependent quantity in Eq. (\ref{FFEqn}) is $\lambda^2 \propto \rho^{-1}$, where $\rho$ is the normalised superfluid density $=n_s / n$. The $T$ dependence of the superfluid density may be derived in the local limit using the following equations appropriate for a two-dimensional cylindrical Fermi surface,\cite{Prozorov}
\begin{eqnarray} 
\rho_{aa/bb}(T)=1-\frac{1}{2\pi k_{\rm B}T}\int_{0}^{2\pi}
\left(
\begin{array}{c}
\cos^2(\phi)\\
\sin^2(\phi)\nonumber \\
\end{array}
\right)\\
\ \ \ \times \int_{0}^{\infty}\cosh^{-2}\Bigl(\frac{\sqrt{\epsilon^2+\Delta_{k}^2(\phi,T)}}{2k_{\rm B}T}\Bigr)d\epsilon d\phi.
\label{superfluid_density}
\end{eqnarray}
Making use of the fourfold rotational symmetry of the crystal (space group:{\it{I}}4/mmm), $\rho_{aa}(T) = \rho_{bb}(T)$, we obtain the following equation for the superfluid density, normalised to its value at $T=0$:
\begin{eqnarray}
\rho(T)\!=\!1\!-\!\frac{1}{4\pi k_{\rm B}T}\!\!\int_{0}^{2\pi}\!\!\!\!\!\int_{0}^{\infty}\!\!\!\!\cosh^{-2}\Bigl(\frac{\sqrt{\epsilon^2\!+\!\Delta_{k}^2(\phi,T)}}{2k_{\rm B}T}\Bigr)d\epsilon d\phi.\nonumber \\
\end{eqnarray}

\begin{figure}[htbp]
\begin{center}
 \includegraphics [width=80mm]{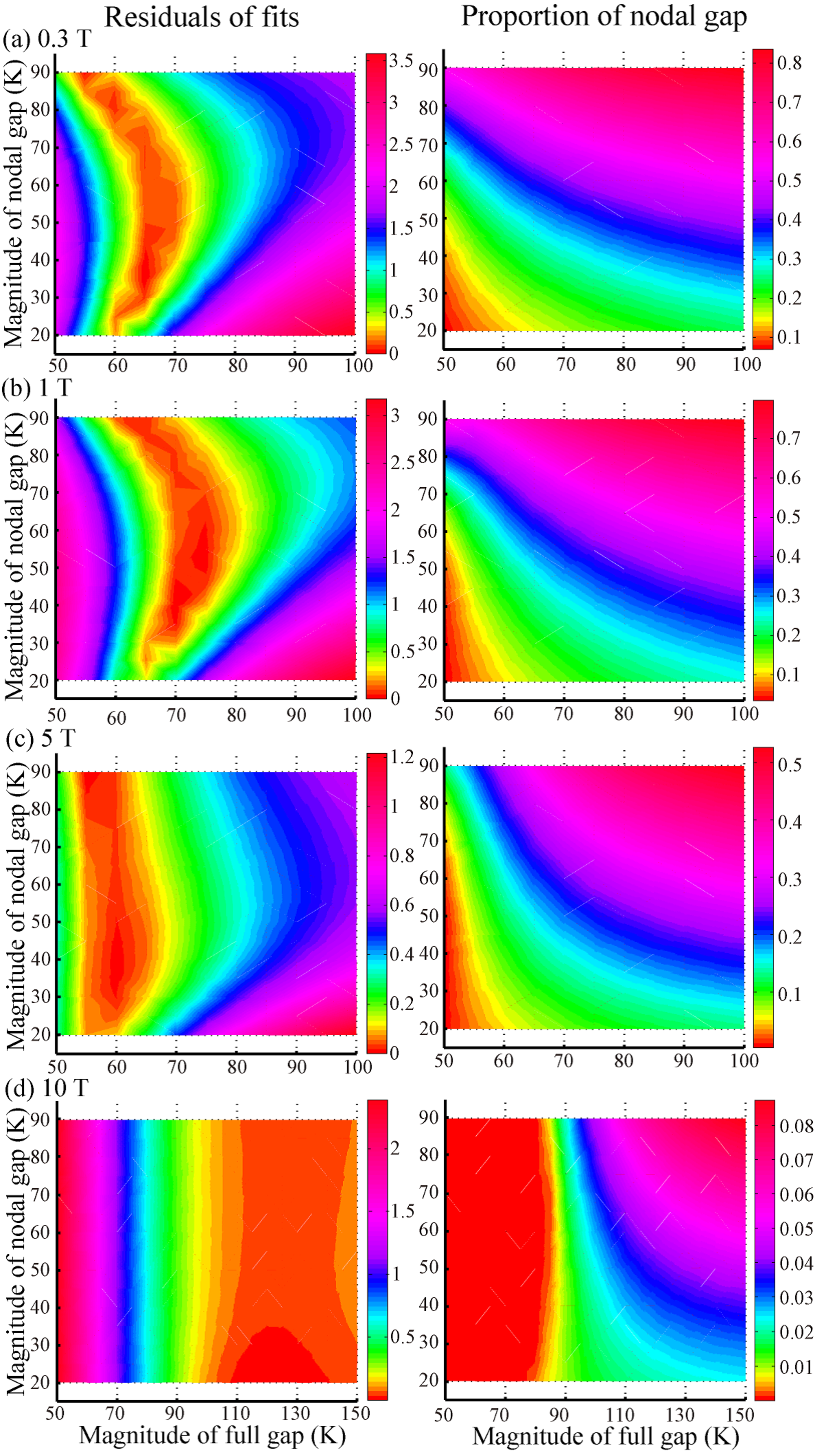}
\end{center}
\caption{(color online) Left panels: Residuals of fits to the temperature dependence of intensity, as the fraction of nodal gap contribution is varied for an array of fixed gap values evaluated at a spacing of 5 K. The horizontal axis is the magnitude of the full gap expressed in Kelvin units; the vertical axis is the magnitude of the maximum of the nodal gap in the same units. The colour scale is the logarithm of residuals divided by their minimum value.  Right panels: proportion of nodal gap contribution given by the same array of fits as above. Note that the scale for 10 T has a smaller range than the others for the nodal contribution to emphasise the disappearance of the nodal gap by 10 T.}
\label{RESID}
\end{figure}

The gap function $\Delta_{k}(\phi,T$) is assumed separable into $T$- and momentum-dependent factors, that is, $\Delta_{k}(\phi,T)=\Delta_{k}(\phi)\Delta_{0}(T)$, where $\Delta_{k}(\phi)$ describes the angular dependence of the gap function, and $\Delta_0(T)$ describes the $T$-dependence. For simple estimation in the weak coupling limit, $\Delta_0(T)$ is given by
\begin{eqnarray}
\Delta_{0}(T)=\Delta_{0}(0)\tanh \Big(1.78\sqrt{\Big(\frac{T_{\rm c}}{T}-1\Big)}\Big),
\label{DeltaWC}
\end{eqnarray}
where $\Delta_{0}(0)$ is the zero-temperature magnitude of the gap. We allow for field effects on the gap by using $T_{\textrm{c2}}(B)$ for $T_{\textrm{c}}$, with values for  $T_{\textrm{c2}}(B)$ taken from Fig.2(e) of our paper. We describe a full gap with $\Delta_{k}(\phi) = 1$, whilst a nodal gap may be represented by $\Delta_{k}(\phi) = \cos(2\phi)$. This is a $d$-wave structure but may be used to model a more general nodal gap~\cite{Furukawa}. 
We found that the data could only be fitted by a \emph {mixture} of nodal and full gap behaviour. To avoid an unreasonably large number of fitting parameters, we restricted the fitting to a simple weighted sum of one nodal and one full gap contribution to the temperature-dependence of the superfluid density. This was used to calculate the temperature-dependence of $\lambda$. The variation in normalised diffracted intensity as a function of temperature was then calculated using the extended London model (Eq. (\ref{FFEqn})) i.e. including core overlap effects representing the finite value of $B/B_{\textrm{c2}}(T)$. For this, the absolute values of $\lambda$ and  $\xi$ at base temperature were taken from our extended London model fit to the field dependence of the vortex lattice form factor. To calculate the temperature dependence of $\xi$, we used Eq. (\ref{DeltaWC}) for the $T$-dependence of the gap and the inverse relationship between $\xi$ and $\Delta$ to give:
\begin{equation}
\xi(T) = \xi(0) \big[ \tanh (1.78\sqrt{(T_{\rm c}/T-1}) \big]^{-1}.
\end{equation} 
This temperature-dependence had a minor influence in the fitting. We found that the residuals of a fit had a broad and shallow minimum as a function of the variables, which were the gap magnitudes and the proportion of full and nodal gaps. This led to problems finding a true minimum at realistic values of parameters. To circumvent this, we fixed the gap magnitudes at pairs of different values and allowed the proportion of full/nodal gaps to vary for each pair of gap magnitudes. This produced a `matrix' of fit results; the residuals and the fitted fraction of nodal gap are represented in Fig.~\ref{RESID}. In these fits, the data were normalised to the extrapolated intensity at $0$ K. 

Because of the broad minima represented in Fig.~\ref{RESID}, minor errors in one or two points can push the absolute minimum of the residuals to unrealistic values, so it is necessary for us to make some reasonable assumptions about the implications of the results. We note that all fits up to $B=$ 5 T indicate the presence of a nodal gap. A field-independent nodal gap value with magnitude of $30$ K is reasonable, since it intersects the fit minimum for all fields up to and including $5$ T. Secondly, a field-independent full gap value of 65 K  is in the `middle ground' for the $0.3$ T to $5$ T fits, and avoids non-monotonic variation in the gap magnitude as a function of field. This leads to a proportion of nodal gap of around $0.2$ for all fields up to and including $1$ T, with $5$ T having around $0.1$ for the contribution of the nodal gap. However, the $10$ T data are consistent with zero nodal contribution, and a rather larger magnitude of the full gap. Whilst it is possible to find lower fit residuals for our data, doing so results in a non systematic variation of gap values, and the broad minimum in all fits indicates that the data does not have enough resolution to demand absolute values in this way. As will be seen in Fig. \ref{IT}, the chosen values of parameters give very acceptable fits to our measured temperature dependences.

We therefore have a picture of almost field-independent behaviour at lower fields, with a significant change in gap structure at higher fields as the nodal proportion begins to decrease by $5$ T and is zero or nearly so at $10$ T. This supports the picture of a mix of full and nodal gaps on the various Fermi surface sheets of BFAP, with the smaller nodal gap being suppressed with increasing fields. This may represent a true disappearance of the nodal contribution to the gap, or the increasing importance of non-local effects at high field, since these would also suppress the linear temperature dependence at low temperature\cite{Amin1,Amin2}. The apparent large value of the gap at 10 T may also reflect nonlocal effects.

From the ARPES experiment\cite{Zhang}, horizontal nodes around a hole Fermi surface near the Z point were observed at zero field, together with nonzero gaps on other Fermi surfaces. Our results for the $T$-dependence of the scaled intensity at 0.3 T, along with our evidence at higher fields against a $d$-wave gap, give strong support for the existence of a horizontal node in the gap structure in BFAP.

\section{Discussion}

We first focus our discussion on the contribution these results make to the debate on the gap structure in the pnictides. There has been significant debate on this subject since the discovery of the iron-based superconductors, with strong evidence for both nodal \cite{Furukawa, Hashimoto, Nakai, Zhang} and nodeless \cite{Zhang2, Hashimoto2, Ding} gap structures. The diffracted intensity from a vortex lattice is related to the superfluid density, making the temperature dependence of the intensity a strong indication of the presence, or lack of, nodes in the gap structure of the superconductor in question. Our investigation into the temperature dependence of the diffracted intensity, shown in Fig. \ref{IT}, presents varying answers to the question of gap structure as a function of field. For lower applied fields, in this case $0.3$ T to $5$ T, an admixture of a fully gapped $s$-wave state and a nodal state present the best fit to the data, whilst at the higher field of 10 T the fully gapped state with no other contribution gives the best description. This can be understood in the context of a multi-band superconductor, with at least one band possessing one or more nodes whilst also having at least one fully gapped Fermi surface sheet. In this case, at lower applied fields all bands are contributing to the superfluid density, leading to indications of nodal behaviour at lower temperatures. At higher applied fields, it appears that the nodal gap has been suppressed by the field, leaving only the full gap, or gaps, contributing to the Cooper pairs. This leads to the $s$-wave like behaviour at lower temperatures and higher fields, where the scattered intensity flattens below $5$ to $10$ K as $k_{\rm B}T$ falls below the gap value, as is clearly seen in Fig. \ref{IT}(d).

Whilst the temperature dependence of the superfluid density is a clear indication of the presence of nodes, it does not necessarily give a direct indication as to the symmetry of the gaps involved.  The structure of the vortex lattice, however, is very sensitive to the gap structure as the applied field and temperature is varied throughout the superconducting mixed state. It is common for superconductor with $d$-wave symmetry to have hexagonal to square VL transitions, such as for CeCoIn$_5$\cite{Bianchi,Kawamura} and YBa$_2$Cu$_3$O$_7$ (YBCO)\cite{White}. Whilst a square scattering pattern is observed for $1$ T, the magnitude of the reciprocal space scattering vector corresponds to a hexagonal lattice and weak diffraction spots are also observed on the horizontal and vertical axes, leading us to believe that this is caused by the adding of multiple diffraction spots from differently oriented domains. The merging of these spots is assisted by weak vortex pinning which is overcome by the inter-vortex interactions between $1$ and $3$ T. Hence these spots give the illusion of a square structure when there is none. The absence of structural transitions in the VL at the highest fields leads us to conclude that the gap structure cannot be $d$-wave. 

Our results suggest that this multi-band superconductor possesses at least one full gap and one nodal gap with circular symmetry. Given that the Brillouin zone possess five Fermi surface sheets, comprising three hole sheets near the centre and two electron sheets at the corners\cite{Zhang}, a mixture of nodal and full gaps is certainly possible. By symmetry, the node structure would therefore take the form of an `accidental' circular line node, similar to that proposed for KFA.\cite{Furukawa} This is in agreement with recent ARPES results on BaFe$_{2}$(As$_{0.7}$P$_{0.3}$)$_2$.\cite{Zhang} Here, they find all the electron sheets of the Fermi surface to be fully gapped, with a pair of circular line nodes occurring in the gap of one of the hole sheets near the Z point of the Brillouin zone. These nodes are caused by a change in sign of the gap function in the region of the Z point. The existence of horizontal line nodes is in apparent disagreement, however, with older ARPES results~\cite{Shimojima}. The newer study notes in their discussion that the older results are not investigating the region of the Brillouin zone where nodes are found, near the Z point, but are in fact investigating the region close to $k_{z} = 0.34 \pi$, and as such would not see evidence of nodes regardless of their presence. Given that we find clear evidence of nodal behaviour in the temperature dependence of the superfluid density in our bulk measurements, we would support these conclusions. Signatures of horizontal line nodes are also found in penetration depth\cite{Hashimoto}, thermal conductivity~\cite{Yamashita}, and NMR results\cite{Nakai}. If horizontal nodes are present, they are a clear sign that a purely two-dimensional model of superconductivity in these materials is insufficient.


These results also give further support to the observation of the existence of nodes in the gaps of pnictides when the distance between the iron and the pnictogen is less than $1.33$ $\rm\AA$.\cite{Kuroki} It is expected that the change between nodeless to nodal states will mark a significant reduction in $T_{\rm c}$. However, BFAP shares a reasonably similar $T_{\rm c}$ with its cousin optimally-doped Ba$_{1-x}$K$_{x}$Fe$_2$As$_2$ ($T_{\rm c}$ $\sim$ 36 K), which has been found to have a nodeless gap structure\cite{Hashimoto2,Evtushinsky,Ding}. This indicates that $T_{\rm c}$ depends on not a nodal gap Fermi surface but any full gap Fermi surface.

\section{Summary}
In summary, we observed the VL in BaFe$_2$(As$_{0.67}$P$_{0.33}$)$_2$ single crystals by SANS measurements over a wide range of applied fields. We found that a nearly regular hexagonal VL is present from 1 T to 16 T; the upper field is a record for observation of the VL by this technique in the pnictides. The absence of a high-field square VL strongly argues against $d$-wave pairing in this material. However, the temperature-dependence of the VL signal  confirms the presence of (non $d$-wave) nodes in the superconducting gap structure at 5 T and below. The nodal effects are suppressed at high fields. At low fields, a VL reorientation transition was observed between 1 T and 3 T, with the VL orientation changing by $45^{\circ}$. However, below 1 T, the VL structure is strongly affected by pinning, and it is likely that the VL reorientation is due to pinning to defects aligned with the crystal structure, rather than being intrinsic.

\section*{ACKNOWLEDGMENTS}
R.M-I. is supported by a JSPS Research Fellowship for Young Scientists. H.K-F. was supported by the Sumitomo Foundation, and Nippon Sheet Glass Company, and Grant-in-Aid for Scientific Research on Innovative Areas (Heavy Electrons) (Grant No. 20102006) of the Ministry of Education, Culture, Sports, Science, and Technology (MEXT), Japan. We are grateful for support and allocated beam time on D11 \& D33 at the ILL, France, on SANS-I at SINQ, PSI, Switzerland, and on CG2 at ORNL, USA. The experiment using the CG2 SANS spectrometer at ORNL was supported by the US-Japan Cooperative Program on Neutron Scattering. We are also grateful to the Swiss NCCR and its program MaNEP and the UK EPSRC.

\end{document}